\documentclass[9pt,conference]{IEEEtran}
\IEEEoverridecommandlockouts

\renewcommand{\baselinestretch}{1.02}

\usepackage{cite}
\usepackage{amsmath,amssymb,amsfonts}
\usepackage{algorithmic}
\usepackage{graphicx}
\usepackage[labelformat=simple]{subcaption}

\usepackage{textcomp}
\usepackage{xcolor}
\def\BibTeX{{\rm B\kern-.05em{\sc i\kern-.025em b}\kern-.08em
    T\kern-.1667em\lower.7ex\hbox{E}\kern-.125emX}}

\begin{document}

\title{
HPU: High-Bandwidth Processing Unit for Scalable, Cost-effective LLM Inference via GPU Co-processing
}

\author{
\IEEEauthorblockN{
\parbox{0.67\linewidth}{\centering
\Large
Myunghyun Rhee, Joonseop Sim, Taeyoung Ahn, Seungyong Lee, Daegun Yoon, Euiseok Kim, Kyoung Park, Youngpyo Joo, Hoshik Kim
}}
\IEEEauthorblockA{
\textit{Memory Systems Research, SK hynix Inc.} \\
\parbox{0.55\linewidth}{\centering
\{myunghyun.rhee, joonseop.sim, taeyoung.ahn, seungyong.lee, daegun.yoon, euiseok.kim, kyoung.park, youngpyo.joo, hoshik.kim\}@sk.com
}}

}

\maketitle

\begin{abstract}
The attention layer, a core component of Transformer-based LLMs, brings out inefficiencies in current GPU systems due to its low operational intensity and the substantial memory requirements of KV caches. We propose a High-bandwidth Processing Unit (HPU), a memory-intensive co-processor that enhances GPU resource utilization during large-batched LLM inference. By offloading memory-bound operations, the HPU allows the GPU to focus on compute-intensive tasks, increasing overall efficiency. Also, the HPU, as an add-on card, scales out to accommodate surging memory demands driven by large batch sizes and extended sequence lengths.
In this paper, we show the HPU prototype implemented with PCIe-based FPGA cards mounted on a GPU system. Our novel GPU-HPU heterogeneous system demonstrates up to 4.1$\times$ performance gains and 4.6$\times$ energy efficiency improvements over a GPU-only system, providing scalability without increasing the number of GPUs.
\end{abstract}


\begin{figure*}[!h]
\centering
    \begin{subfigure}[t]{0.3\linewidth}
      \includegraphics[width=\textwidth]{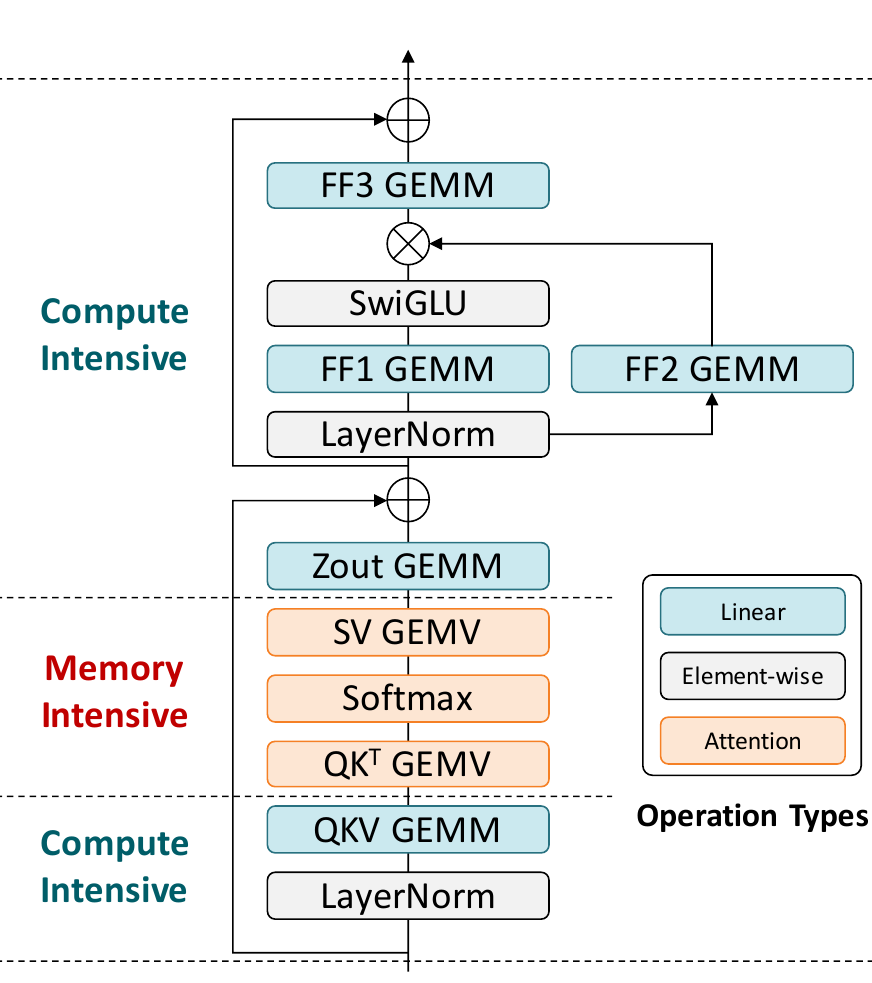} 
      \caption{}
      \label{fig:llama2_arch}
    \end{subfigure}
    \hspace{2mm}
    \begin{subfigure}[t]{0.3\linewidth}
        \includegraphics[width=\textwidth]{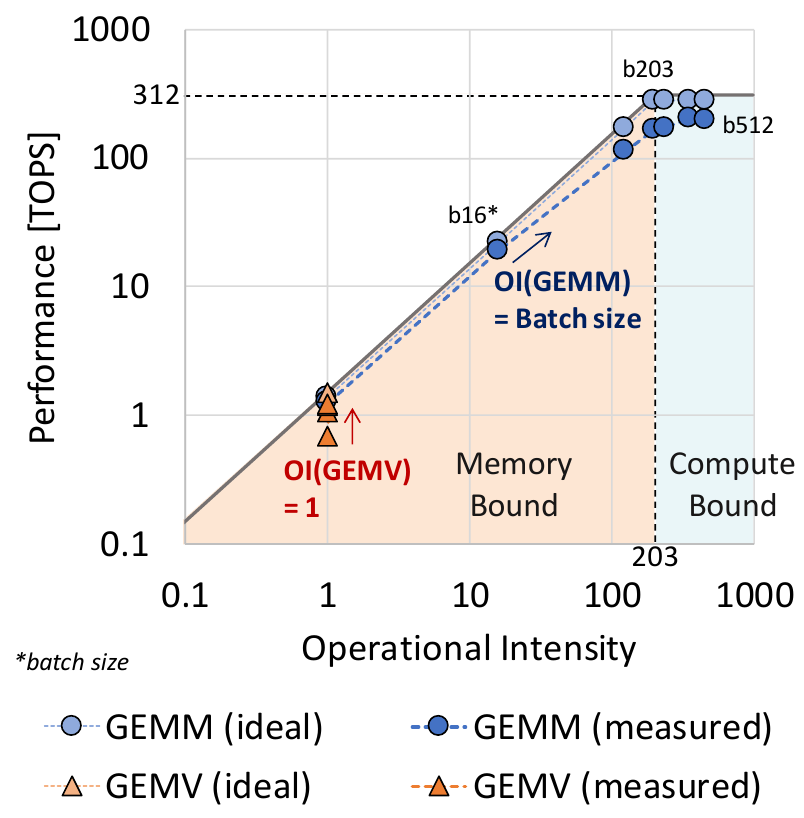} 
        \caption{}
        \label{fig:roofline}
    \end{subfigure}
    \hspace{2mm}
    \begin{subfigure}[t]{0.3\linewidth}
        \includegraphics[width=\textwidth]{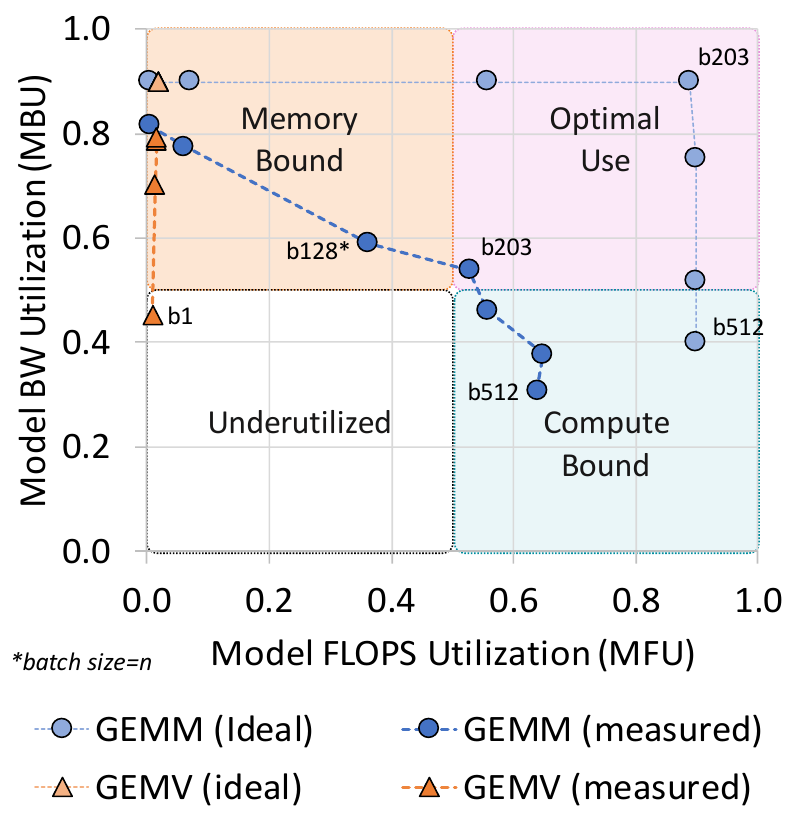} 
        \caption{}
        \label{fig:mfu_mbu}
    \end{subfigure}
\caption{LLM inference characteristics.
(a) A single Transformer layer architecture of Llama 2. In multi-batch scenarios, computations are composed of GEMM-based linear layers and GEMV-based attention layers.
(b) Roofline of the A100 GPU and the trend of GEMM and GEMV characteristics as batch size increases.
(c) GPU utilization from the perspective of MFU (Model FLOPS Utilization) and MBU (Model Bandwidth Utilization).}
\label{fig:gpu_util}
\end{figure*}

\section{Introduction}
\label{sec:intro}

\label{sec:motive}

The rapid advancements in large language models (LLMs), such as GPT~\cite{gpt} and Llama~\cite{Llama}, have redefined natural language processing (NLP) by achieving exceptional performance across diverse applications. These models leverage the Transformer architecture, which surpasses earlier approaches like RNNs~\cite{grossberg2013recurrent} and LSTMs~\cite{graves2012long} by capturing long-range dependencies and enabling efficient parallel processing~\cite{vaswani2017attention}. Despite their transformative capabilities, LLMs also pose significant computational challenges, particularly during inference, where the attention layers dominate both memory and compute workloads.

A key component of Transformer models is the attention layer, which computes the importance of tokens relative to one another in an input sequence. However, this layer exhibits low operational intensity (Ops/Byte) due to the significant data access required to compute attention scores, especially as sequence lengths and batch sizes increase~\cite{shen2021efficient}. Unlike compute-intensive fully connected (FC) layers, attention layers are memory-bound, relying heavily on General Matrix-Vector Multiplication (GEMV) operations~\cite{park2024attacc}. This mismatch leads to \textit{under-utilization of GPU computation resources}, constrained by memory bandwidth limitations~\cite{he2024fastdecode}.

A major contributor to these inefficiencies is the key-value (KV) cache, a critical component for optimizing autoregressive token generation. The KV cache stores intermediate outputs from the query (Q), key (K), and value (V) matrices. During inference, the KV cache eliminates the need to recompute these matrices for previously processed tokens, instead reusing the stored values to calculate attention scores for new tokens. This approach significantly reduces computational complexity, lowering it from quadratic (due to repeated matrix multiplications) to linear with respect to sequence length. 

However, the KV caching mechanism comes with substantial trade-offs. As model sizes grow, so do the dimensions of the K and V matrices, resulting in exponential increases in memory capacity and bandwidth requirements. These demands scale further with batch size and sequence length~\cite{shen2021efficient}, leading to severe inefficiencies in memory-bound operations. In large-batch or long-sequence inference scenarios, the bandwidth saturation and high memory footprint of the KV cache often become the primary bottlenecks, substantially increasing computational costs and limiting throughput.

To address these challenges, we propose the High-Bandwidth Processing Unit (HPU), a memory-intensive yet lightweight co-processor designed to complement GPUs during LLM inference. 
The HPU offloads the memory-bound attention computations, allowing GPUs to focus on compute-intensive tasks and thereby maximizing overall resource utilization.
Additionally, designed as an add-on card type form factor, the HPU provides a scalable and cost-effective solution to meet the growing KV cache demands without requiring additional GPUs.
This design not only alleviates GPU memory bandwidth bottlenecks but also enables scalable LLM inference for larger batch sizes and longer sequence lengths without increasing the number of GPUs.

In our evaluation, we show that HPU provides a practical and cost-effective solution for scaling LLM inference by handling the KV cache demands combined with its specialized attention layer acceleration.
Our proposed GPU-HPU heterogeneous system achieves up to  4.1$\times$ throughput and 4.6$\times$ energy efficiency improvements compared to GPU-only systems.

The main contributions of this paper are as follows:
\begin{itemize}
    \item We propose the High-Bandwidth Processing Unit (HPU): a memory-intensive yet lightweight co-processor for LLM serving along with GPUs.
    \item We propose a novel GPU-HPU heterogeneous system architecture that maximizes system throughput and GPU resource utilization by offloading the \textit{attention} layers with low operational intensity to HPU.    
    \item We implemented the HPU prototype as a PCIe card and end-to-end GPU-HPU heterogeneous system for LLM inference. To the best of our knowledge, HPU is the first real-world prototype of an HBM-based attention accelerator to complement GPUs.
    \item HPUs effectively scale and handle substantial memory capacity and bandwidth requirements of KV cache in batched inference requests by providing \textit{attention} acceleration cards, thus satisfying both scalability and energy-efficiency.
\end{itemize}

\section{Related Work}
\label{sec:related_work}

Efforts to enhance GPU utilization for running LLM applications include strategies such as optimizing memory usage~\cite{ainslie2023gqa} or dynamically adjusting batch size~\cite{yu2022orca}.
They introduced ways to alleviate memory constraints by sharing key/value data across multiple queries or to improve GPU resource efficiency by overlapping the summarization and generation stages. Despite these advancements, their operational intensities remain low, around 8-10, falling significantly short of the high FLOPS capabilities (tens to hundreds) of modern GPU systems~\cite{a100,h100}.

Several studies improved the efficiency of GPUs by configuring heterogeneous systems. The work in~\cite{choi2023unleashing,park2024attacc,heo2024neupims} used HBM-based Processing in Memory (PIM) architecture that exploits high internal bandwidth to process memory-intensive attention layers while GPU/NPU is in charge of compute-intensive fully-connected layers.
PIM technology is suitable for processing applications that have extremely low operational intensity of 1 Ops/Byte or less by fully utilizing the in-memory bandwidth. However, for applications with larger than a few Ops/byte, the PIM device has spatial and cost constraints in implementing additional logic. The work in~\cite{yun2024duplex} used the HBM's base die to process operations with operational intensities of 1$-$10.
However, since it should share the memory space between the GPU and the computational functions in a base die, unnecessary data copying due to differences in data placement between the two and a complex MMU function to share virtual addresses are substantial challenges.

Our HPU has advantages over the mentioned state-of-the-art works in the following aspects: 
1) spatial flexibility sufficient to design a few OPS/Byte level computational functions with add-on type FPGA cards (possibly SoC in the future), 
2) outstanding system scalability in operating large batched LLM inference, 
and 3) implementing the real system for valid performance evaluation
(details in Section~\ref{sec:eval}).

\begin{figure*}[t]
\centering 
\includegraphics[width=0.95\linewidth]{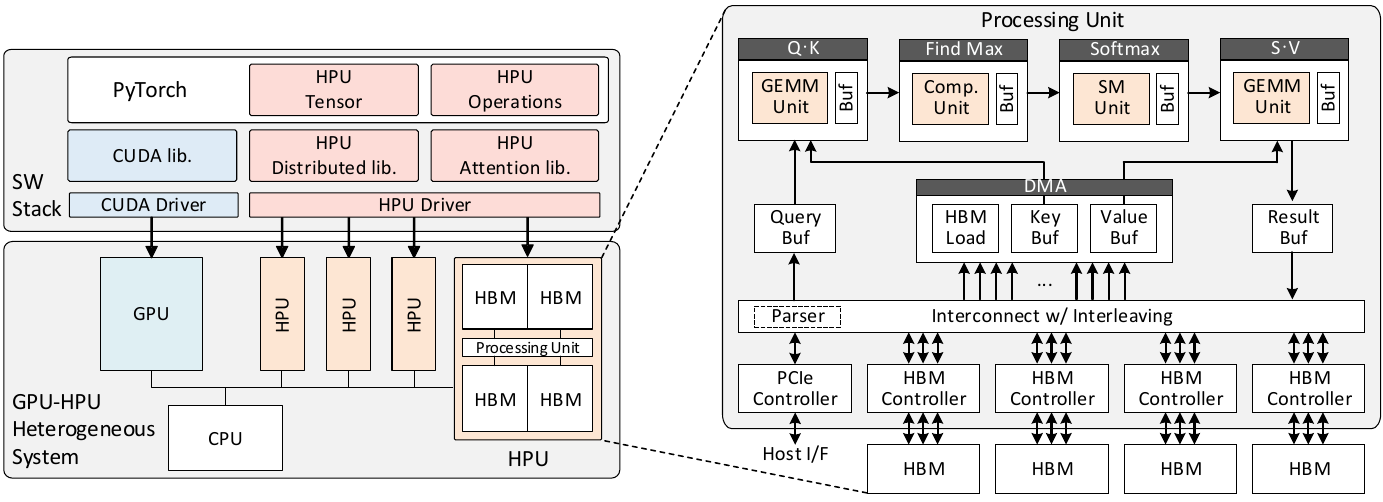}
\begin{minipage}{0.47\textwidth}
    \centering
    \subcaption{GPU-HPU heterogeneous system with SW stack.}
    \label{fig:gpu_hpu_hetero}
\end{minipage}%
\hfill
\begin{minipage}{0.5\textwidth}
    \centering
    \subcaption{Proposed HPU architecture.} \label{fig:hpu_arch}\label{fig:hpu_proto_arch}
\end{minipage}
\caption{Overview of the proposed HPU solution.}
\label{fig:hpu_arch_overview}
\end{figure*}

\section{GPU Profiling}
\label{sec:profiling}
To evaluate the impact of batch size on GPU utilization, we profiled GPU kernels while running Llama 7B~\cite{touvron2023llama}.
The structure of a single Transformer block is shown in Fig.~\ref{fig:llama2_arch}. For our analysis, 
we selected the QK GEMV block within the attention layer and the third FC layer as representative kernels for GEMV and GEMM, respectively.
We used an A100 PCIe GPU, which offers 1.55TB/s of memory bandwidth, 40GB of memory capacity, 
and 312 TFLOPS (FP16, no sparsity) of computing performance. 
The performance-to-bandwidth ratio of the A100 GPU is approximately 203, 
meaning that applications with a similar operational intensity (OI) are expected to perform optimally. 
We increased the batch size progressively from 1 to 512 and profiled the CUDA kernels using NVIDIA NsightSystems~\cite{nsys} and NsightCompute~\cite{nc}.

Fig.~\ref{fig:roofline} presents the profiling results using a Roofline analysis. The GEMV operation, which dominates the attention layer, shows no change in OI as batch size increases, meaning that it remains memory-bound. However, the GEMM operation, which makes up the majority of the linear computations, exhibits that OI increases as the batch size changes.
The theoretical prediction (GEMM ideal) aligned well with the measured results (GEMM measured), demonstrating that increasing the batch size effectively improves GPU performance. Notably, we observed that performance saturation occurs at batch size 203, where the performance characteristics of the A100 GPU shift from memory-bound to compute-bound.

Fig.~\ref{fig:mfu_mbu} shows the comparison of Model FLOPS Utilization (MFU)~\cite{chowdhery2024palm} and Model Bandwidth Utilization (MBU)~\cite{mbu} between GEMV and GEMM operations.
Memory-bound operations typically embody high MBU but low MFU, while compute-bound operations show the opposite.
When both MBU and MFU are high, 
it indicates an ideal scenario where all GPU resources are maximally utilized.
In GEMV operations, memory-bound behavior dominates, keeping MFU consistently low across all batch sizes. However, as batch size increases, GEMV's MBU improves, nearing its theoretical maximum of 90\%.
As discussed earlier, at the 203 batch, where the workload’s OI aligns with the A100 GPU's performance/bandwidth ratio, both MBU and MFU are efficiently balanced. When the batch size exceeds 256, GEMM fully transitioned into a compute-bound regime, where MFU continued to increase while MBU declined, indicating inefficient memory usage.

To summarize the key findings obtained from the profiling results to optimize GPU utilization, (1) memory-bound operations should be minimized, and (2) the application’s operational intensity should match the GPU’s performance-to-bandwidth ratio. In the context of LLM inference, attention layers remain memory-bound regardless of batch size, making it difficult to use GPU resources efficiently. On the other hand, GEMM operations, which see an increase in OI as batch size grows, require sufficiently large batch sizes to match the GPU’s characteristics for optimal performance.

\section{Proposed Architecture}
\label{sec:hpu}
\subsection{High-Bandwidth Processing Unit (HPU)}
\label{sec:hpu_arch}
We propose the High-Bandwidth Processing Unit (HPU), to address the challenges of LLM inference as outlined in Section~\ref{sec:related_work}. The HPU is a cost-effective, lightweight attention accelerator designed to handle the generation-stage attention layer operations in LLM inference. Rather than executing the full end-to-end LLM inference pipeline, the HPU is designed to offload specific memory-intensive operations, optimizing overall system efficiency. As shown in Fig.~\ref{fig:hpu_arch}, the HPU is equipped with four stacks of HBM3e 12-Hi memory for high bandwidth, providing a total bandwidth of 4.9TB/s and 144GB of capacity, with a pin speed of 9.6 Gbps~\cite{hynix_hbm3e12}. This configuration surpasses the memory bandwidth offered by the latest H100 NVL GPU, emphasizing the HPU’s specialization in memory performance. In contrast, its compute resources are limited only for executing scaled-dot-product attention. The HPU processing units feature a narrow GEMM engine, optimized for Grouped-Query Attention (GQA)~\cite{ainslie2023gqa} delivering 39.3 TFLOPS (FP16), which is 8 times higher than its bandwidth. This design allows the HPU to accelerate GQA without performance degradation for the number of heads per each group up to 8. Additionally, the HPU includes a softmax unit, enabling it to compute attention results based on the query, key, and value tensors provided by the GPU. A detailed HW architecture description is elaborated in Section~\ref{sec:prototype_hw}

\subsection{GPU-HPU Heterogeneous System}
\label{sec:hpu_hetero}
The HPU is designed specifically to accelerate the part of LLM inference, meaning it must collaborate with the GPU to perform the complete end-to-end inference process. This design is particularly beneficial in scenarios with large batch sizes and linear parts become more compute-intensive, which increases GPU utilization. Fig.~\ref{fig:gpu_hpu_hetero} shows the GPU-HPU heterogeneous system and SW stack. 

By offloading the memory-bound attention layer to the HPU, the system can maintain high utilization for both compute and memory resources. It significantly improves overall throughput (detailed in Section~\ref{sec:eval_throughput}), while HPU's additional memory capacity allows larger batch sizes and longer sequence lengths. It allows the HPU to enhance the scalability of LLM inference systems, enabling efficient use of existing GPU infrastructure without requiring substantial hardware changes (detailed in Section~\ref{sec:eval_scal} and~\ref{sec:eval_util}). 

As each device handles only a part of the LLM inference, requiring frequent data transfers between devices, the network capability of the HPU is critical. However, since the transferred data consists only of generated query, key, and vector tensors, the data size remains negligible compared to weight parameters or KV cache. Our calculations and experimental results indicate that the PCIe network bandwidth between GPU and HPU is not the bottleneck point to handle these transfers, 
as we will discuss further in Section~\ref{sec:eval_throughput}.

\begin{figure}[t]
\centering
\includegraphics[width=1\linewidth]{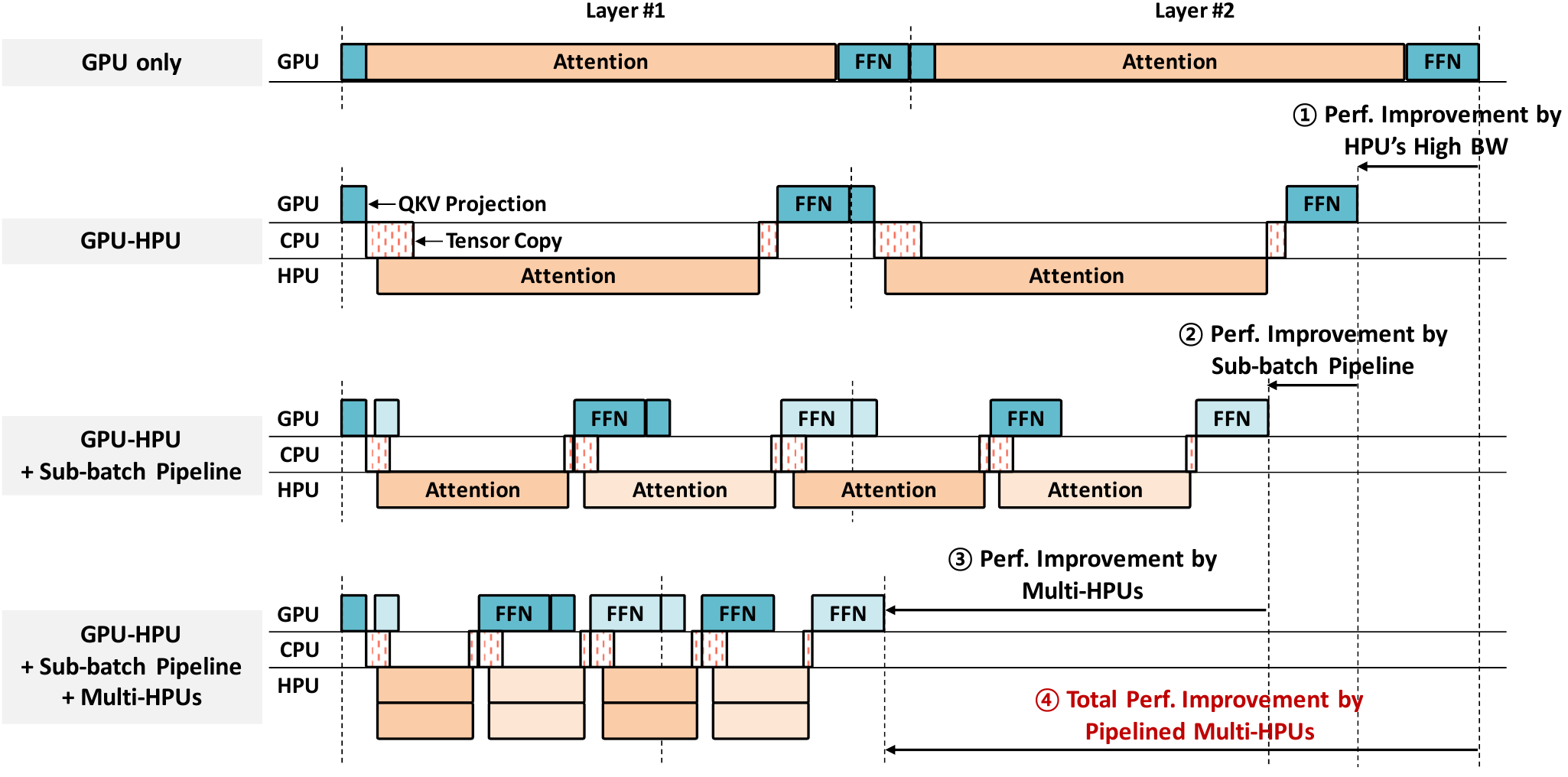}
\caption{GPU-HPU pipelined timing diagram.}
\label{fig:gpu_hpu_timing}
\end{figure}

\subsection{Parallelism}
\label{sec:hpu_parallel}
To maximize the performance and efficiency, we devised parallelization techniques in the HPU design.
The parallelism in HPU can be classified into two types: (1) parallelism between the GPU and HPU, and (2) parallelism across multiple HPUs. Both are necessary to fully leverage the system’s capabilities and ensure high throughput and resource utilization.

For GPU-HPU parallelism, the key is to organize the computation into a pipeline that allows both devices to work concurrently, avoiding idle times on either side. Since the HPU is specialized for attention layer operations and the GPU handles the rest of the inference, the workload needs to be split in a way that both devices can be fully utilized. This is done by dividing the batch into sub-batches and processing them in a staggered fashion. While the HPU processes the attention for one sub-batch, the GPU performs feed-forward computations for another sub-batch. This pipelining technique is similar to the approach used in GPU-only systems for model parallelism. Fig.~\ref{fig:gpu_hpu_timing} shows an overall diagram of parallel processing for the GPU-HPU heterogeneous system.

\begin{figure}[t]
\centering
    \begin{subfigure}[t]{\linewidth}
    \includegraphics[width=\textwidth]{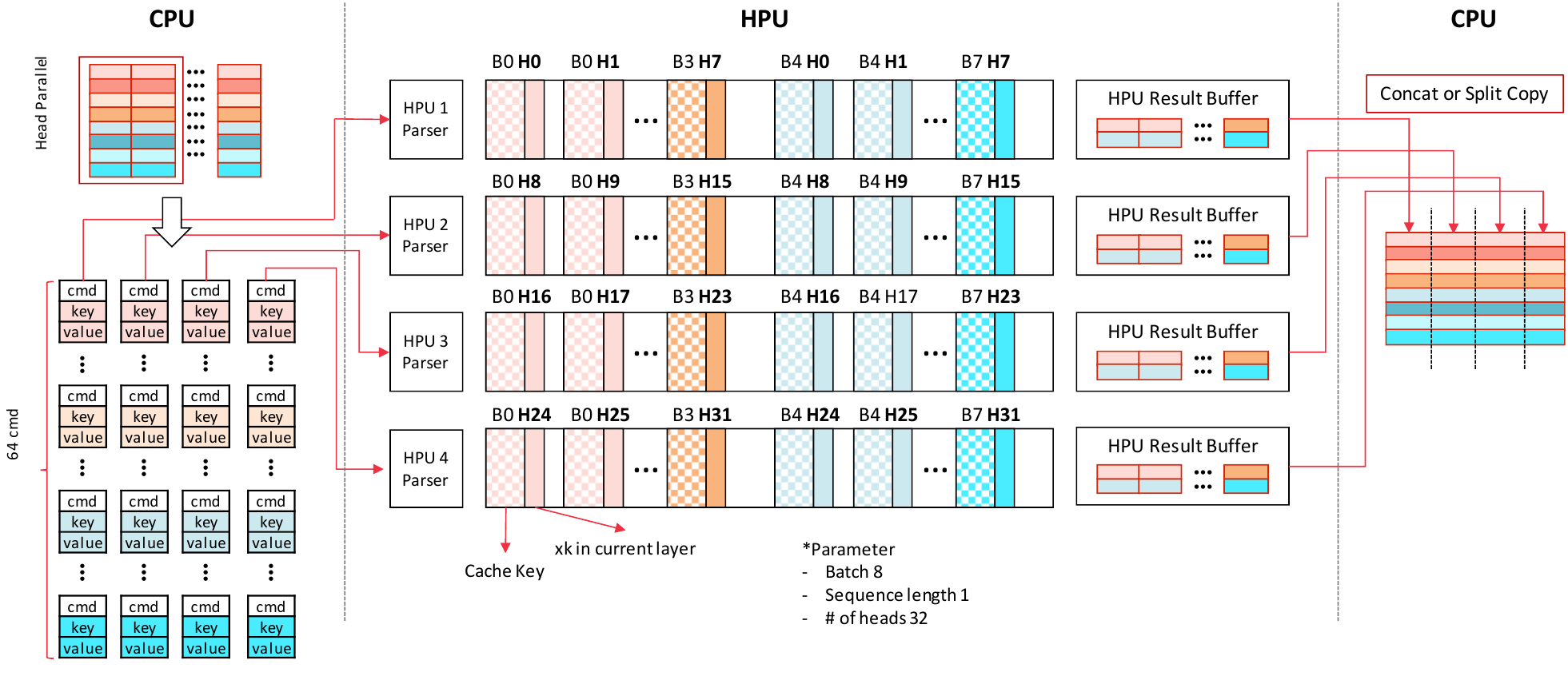} 
    \caption{Head-based parallelism.}
    \end{subfigure}
    \begin{subfigure}[t]{\linewidth}
    \includegraphics[width=\textwidth]{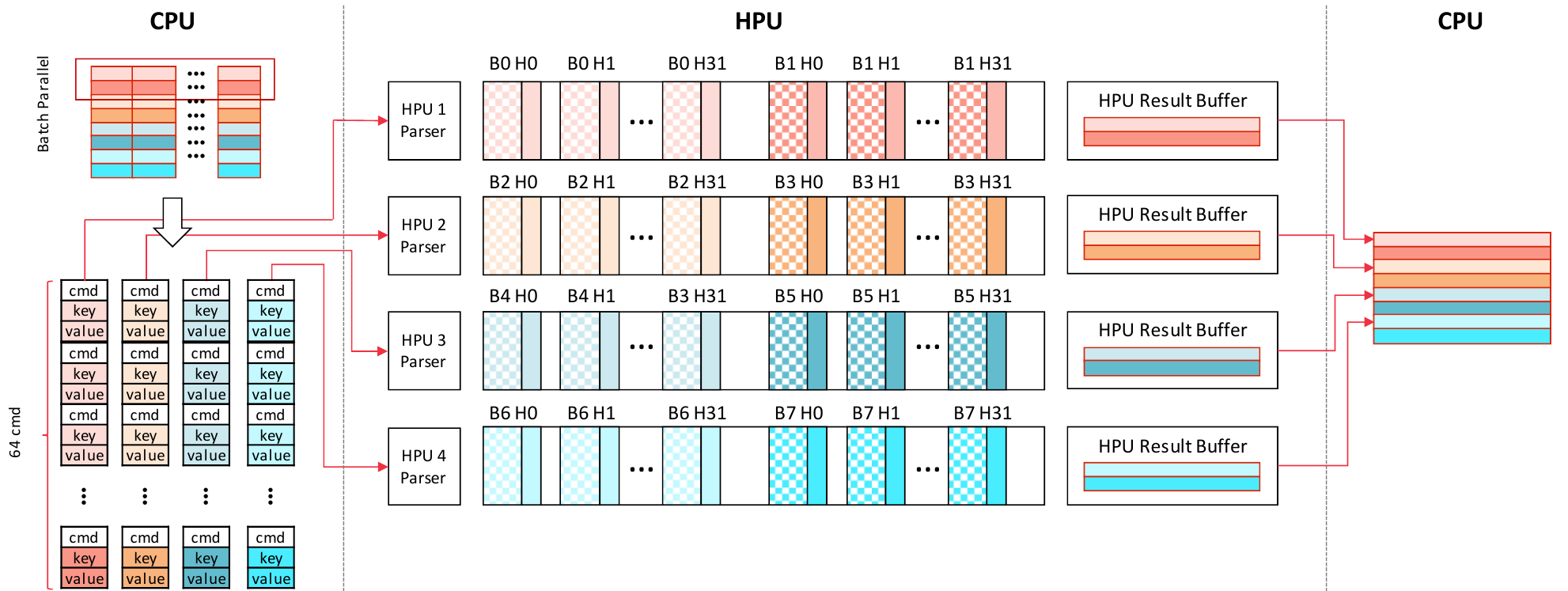} 
    \caption{Batch-based parallelism.}
    \end{subfigure}
\caption{Parallelism and data placement policy to optimize performance in multi-HPU environments.}
\label{fig:multi_hpu_parallel}
\end{figure}

One of the challenges in this approach is that the computation times for the attention layer operations on the HPU and the linear operations on the GPU may not always be balanced. If the HPU finishes its work significantly faster than the GPU, it may remain underutilized, and vice versa. To mitigate this, the system must be tuned by adjusting the batch size and sequence length so that the execution times for attention and linear layer operations are as closely matched as possible. This balance maximizes the overall throughput of the system, ensuring both devices are operating at peak efficiency.

We also enabled a parallelism manner between HPUs that can efficiently distribute the workloads across multiple HPUs as shown in Fig.~\ref{fig:gpu_hpu_timing}. This is particularly beneficial when dealing with very large models or when serving multiple inference requests concurrently. In a multi-HPU setup, each HPU can handle different portions of the attention workload, either by splitting the input across attention heads or distributing a subset batch of the entire batch across multiple devices. Multi-HPU parallel processing enables the system to meet surging inference demands, increase throughput, and ensure scalability.
As shown in Fig.~\ref{fig:multi_hpu_parallel}, we also employ two types of partitioning policies, head-based parallelism and batch-based parallelism. They respectively parallelize the head and batch-wise workloads with the stages of issuing parallel computation, transferring data, and gathering results in a multi-HPU setup.

When applying the parallelism techniques of Fig.~\ref{fig:gpu_hpu_timing}, optimization to avoid communication bottlenecks between the GPU and multiple HPUs is critical. Efficient data partitioning and routing strategies need to be employed to ensure that the relatively small query, key, and value tensors are transferred to the HPUs without significant overhead. Our experiments have shown that PCIe bandwidth overhead is not significant (detailed in Section~\ref{sec:eval_throughput}), though more advanced interconnects like NVLink could enhance scalability and reduce latency in larger deployments.

\begin{table*}[!t]
\centering
\caption{Comparison of GPU, HPU specification}
\label{tab:spec}
    \begin{tabular}{|c|c|c|c|c|} \hline 
        \textbf{Description}& \textbf{L40S} & \textbf{H100 NVL} &  \textbf{HPU} &  \textbf{HPU Prototype}    \\ \hline\hline
        Memory BW (GB/s)    & 864       & 3,900     & 4,900    & 460            \\ \hline 
        Memory Capacity (GB)& 48        & 96        & 144      & 16              \\ \hline 
        Performance
        (FP16 TFLOPS)       & 362.1     & 835.5     & 39.3     & 0.5            \\ \hline 
        Performance/BW      & 419       & 213       & 8        & 1              \\ \hline
        Network             & PCIe 4.0 x8 16GB/s & NVLink 900GB/s& PCIe 5.0 x16 64GB/s&  PCIe 4.0 x8 16GB/s\\ \hline 
        TDP (Watt)          & 350       & 400       & 120       & 150           \\ \hline
        Form Factor         & PCIe FHFL & PCIe FHFL & PCIe FHHL & PCIe FHHL     \\ \hline
    \end{tabular}    
\end{table*}

\section{HPU Prototyping}
\label{sec:prototype}
We implemented the HPU prototype and GPU-HPU heterogeneous system using AMD Alveo U55C cards and a L40S GPU.
This section provides a detailed description of the HW and SW configurations of the HPU prototype.

\begin{figure}[t]
\centering
\includegraphics[width=1\linewidth]{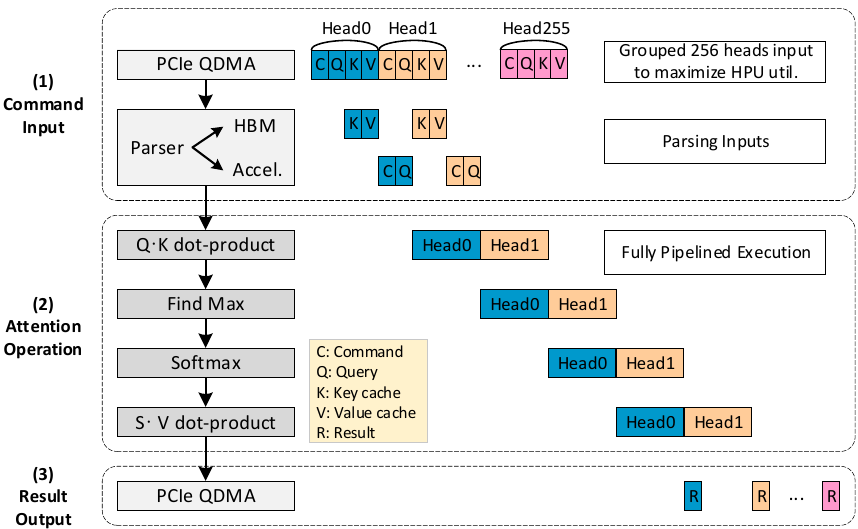} 
\caption{HPU Prototype execution pipeline.}
\label{fig:hpu_proto_pipeline}
\end{figure}

\subsection{Prototype HW Implementation}
\label{sec:prototype_hw}
The architecture of the prototype HW is identical to Fig.~\ref{fig:hpu_proto_arch}, but due to FPGA resource constraints, the prototype is optimized for Multi-Head Attention (MHA). To fully exploit the HBM2 x2 @460GB/s bandwidth of the U55C card, we implemented GEMV units capable of delivering up to 460 GFLOPS. However, we also support Grouped Query Attention (GQA), and if the GEMV unit can be expanded, GQA optimal performance can be implemented. 

The HPU prototype performs scaled dot-product attention at the attention heads level, supporting 128 dimensions with FP16 floating point arithmetic. For the attention layer computation, the system receives commands, along with query, key, and value tensors generated by the GPU, via PCIe QDMA.
The command includes the input sequence length for token generation and the base address of the KV cache. The received command and query are stored in internal buffers, while the key and value tensors are stored in HBM before the attention layer operation is executed.

Due to the nature of DMA, PCIe QDMA suffers from significant throughput degradation when transferring small data sizes, which can result in a performance bottleneck for the HPU. 
To address this issue, we optimized throughput by grouping 256 head inputs into a single chunk and transmitting them in bulk through PCIe QDMA. Upon receiving the chunk, the HPU’s internal parser logic distributes the command and query to the attention accelerator while the key and value tensors are split and stored in HBM, with addresses calculated from the KV cache address and sequence length provided in the command (See Fig.~\ref{fig:hpu_proto_pipeline} (1)).

To maximize HBM access efficiency, the KV cache is interleaved in 64B blocks using a hardware interleaver and stored in HBM. The attention accelerator accesses these blocks in parallel through multi-port access, thereby enhancing memory utilization. Intermediate results between computational blocks are transferred through internal buffers to reduce latency. However, as the size of these intermediate results grows with the sequence length, supporting sequences longer than several tens of thousands of tokens requires offloading these buffers to reserved regions in HBM. Although this may introduce additional latency, the impact is expected to be minimal, as the longer sequence lengths naturally increase the overall token generation time.

As shown in Fig.~\ref{fig:hpu_proto_pipeline}, the HPU pipeline spans from PCIe QDMA input to each computational block, effectively masking network overhead and achieving high throughput during computations. In the case of an input sequence length of 2K, the attention accelerator achieved a high memory utilization of 73\%. As the input sequence length increases, the continuous read operations from the KV cache are expected to increase, further improving memory utilization.

\begin{figure}[t]
\centering
\includegraphics[width=0.8\linewidth]{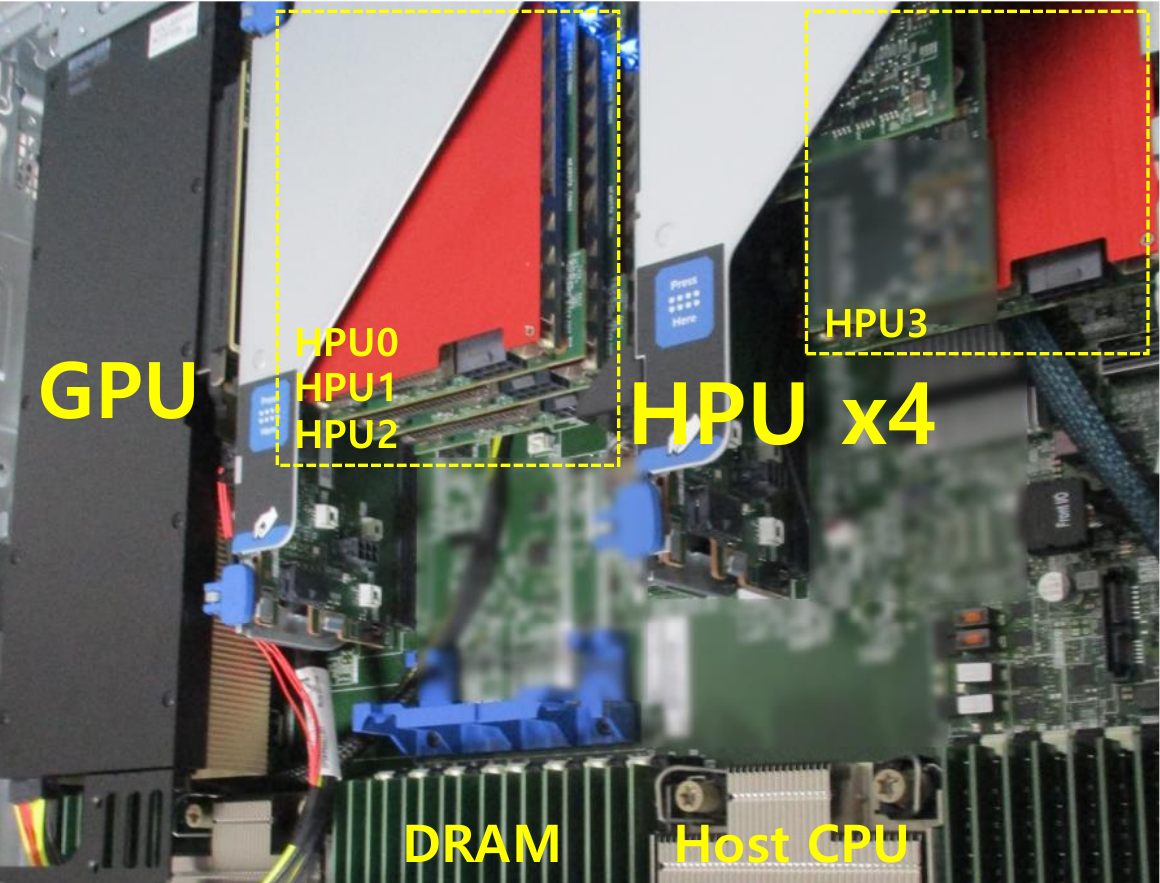} 
\caption{The real-world heterogeneous server system configured with a GPU and 4 HPU prototypes, identical to the system setup shown in Fig.~\ref{fig:hpu_arch_overview}.}
\label{fig:gpu_hpu_proto}
\end{figure}

\subsection{Prototype SW Implementation}
\label{sec:prototype_sw}
We modified the PyTorch-based Llama 2 code and implemented the HPU SW in C++ to offload the scaled dot-product attention computation to the FPGA-based HPU prototype. This SW manages the KV cache and orchestrates the movement of commands and data between the GPU and HPU. 

During the summarization stage, the HPU SW moves the key and value caches into the HPU prototype. These caches are placed contiguously in HBM based on sequence length and organized by the head dimension. This structure is designed to maximize read bandwidth, thereby optimizing HPU performance. In the generation stage, query, key, and value tensors are grouped by head dimension and packaged into a descriptor before being transmitted to the HPU. A descriptor encapsulates metadata such as the KV cache address and sequence length along with the query, key, and value tensors. Due to FPGA board constraints, data cannot be transferred directly between the GPU and HPU prototype, necessitating routing through the CPU. However, as described earlier, optimizing the CPU-HPU pipeline mitigates most of the network overhead.

The HPU supports scaling to multiple units, allowing for different KV cache-loading policies. 
In our implementation, we supported both head-parallel and batch-parallel processing (Fig.~\ref{fig:multi_hpu_parallel}). While both methods result in similar compute performance within the HPU prototype, the process of merging results differs. Head-parallel processing slices vectors across multiple HPU prototypes, which introduce merging overhead on the host side, as results must be combined per batch. In contrast, batch-parallel processing allows contiguous merging of results, resulting in superior performance. Based on this, we applied the batch-parallel method for the experiments presented in this paper.

\begin{figure}[!t]
\centering
\begin{subfigure}[t]{\linewidth}
    \centering
    \includegraphics[width=\linewidth]{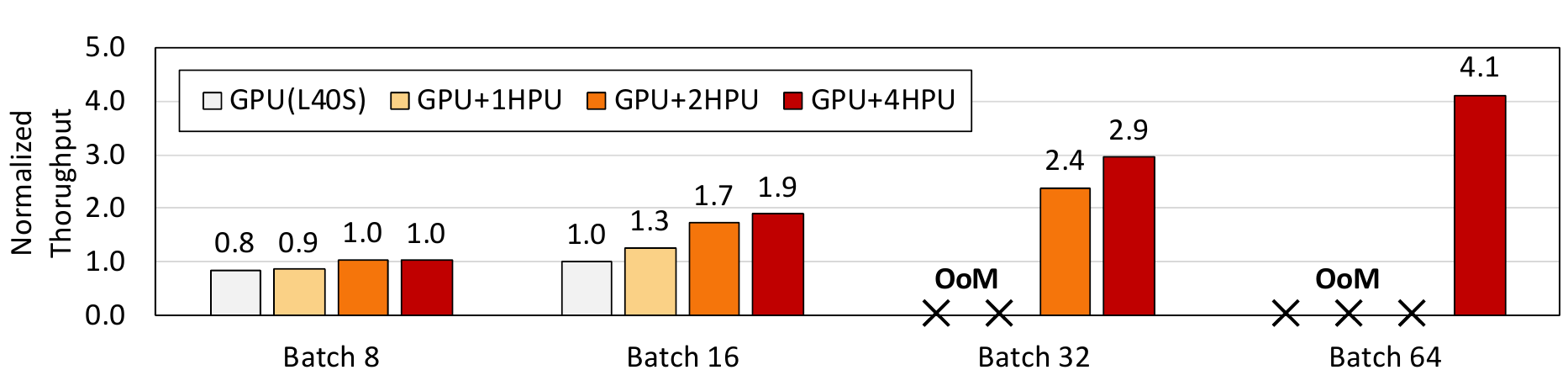}
    \caption{LLM inference performance GPU only system vs. GPU-HPU prototype heterogeneous system.}
    \label{fig:eval_throughput}
\end{subfigure}
\begin{subfigure}[t]{\linewidth}
    \centering
    \vspace{4mm}
    \includegraphics[width=\linewidth]{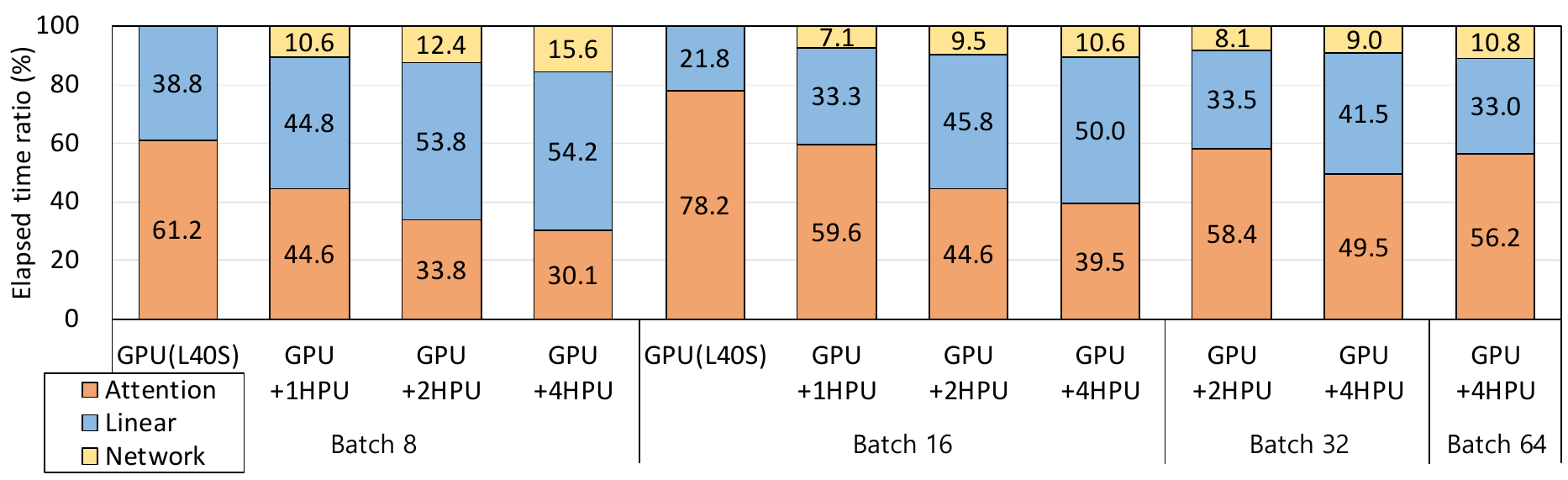}
    \caption{Execution time breakdown of generation stage.}
    \label{fig:eval_time_breakdown}
\end{subfigure}
\label{fig:eval}
\caption{LLM inference throughput and execution time breakdown.}
\end{figure}

\section{Evaluations}
\label{sec:eval}
\subsection{Experimental Setup}
\label{sec:eval_exp_setup}
To demonstrate the validity of the HPU architecture, we performed detailed performance and power measurements in a real-world experimental setup. The GPUs used in the experiments were the NVIDIA L40S and H100 NVL, while the HPU prototype was implemented using an AMD U55C FPGA board. The L40S was employed for both GPU-only and GPU-HPU heterogeneous system experiments, whereas the H100 NVL was used exclusively for GPU-only. As shown in Fig.~\ref{fig:gpu_hpu_proto}, our experimental server could host a L40S and up to 4 HPU prototypes, denoted as 'HPU prototype x4' or '4HPU'. Due to the substantial difference in hardware specifications between the H100 NVL and the HPU prototype, it was not feasible to balance them for a heterogeneous system configuration; thus, we set aside the H100 NVL + HPU prototype system from experiments. However, once the target HPU is fully developed, the H100 NVL is expected to be suitable for collaborative operations.
The hardware specifications for the devices used are detailed in Table~\ref{tab:spec}.

The AI model used for the experiments was Llama 2 7B, with synthetic data designed to optimize throughput measurement. The batch size varied from 1 to 64, and the sequence length was configured as input 1K, output 1K, total 2K. End-to-end execution time was measured across all systems, and throughput was calculated based on these measurements. Power consumption was monitored using APC Rack Automatic Transfer Switch (AP4423A), and the power values were derived from the difference between the system's power consumption during LLM inference and its idle state.

\subsection{Scalability}
\label{sec:eval_scal}
The scalability of the GPU-HPU prototype heterogeneous system was evaluated against a GPU-only system, 
focusing on batch size limitations and throughput. As shown in Fig.~\ref{fig:eval_throughput}, for the Llama 2 7B model with a context length of 2K (input and output 1K each), the GPU-only system was limited to a batch size of 16 due to memory capacity constraints, and attempts to increase the batch size to 32 or more resulted in Out of Memory (OoM) errors. The HPU prototype, with 16GB of memory per unit, could store KV cache and perform attention layer operations for up to 16 batches per unit. By offloading the KV cache and the attention to the HPU prototype, we demonstrated that the system can handle larger batch sizes without increasing the number of GPUs. 

\subsection{Throughput \& Time Breakdown}
\label{sec:eval_throughput}
The key performance metric for LLM inference throughput is the number of generated tokens per second. To simplify the analysis, 
we normalized the tokens/s measurements based on the GPU-only system with 16 batches. When four HPU prototypes were used, increasing the batch size from 16 to 32 and 64 resulted in throughput improvements of 1.9$\times$, 2.9$\times$, and 4.1$\times$, respectively. However, with smaller batch sizes, such as 8, the offloading overhead and insufficient HPU utilization led to minimal performance gains, indicating that the GPU-HPU system is optimized for large-batch processing. Fig.~\ref{fig:eval_time_breakdown} shows the time distribution of generation stage operations during inference, with network overhead from tensor transfers remaining low, around 10\%, even as the batch size increases. This confirms that, as discussed in Section~\ref{sec:hpu_parallel}, the pipelined operations effectively hide network overhead.

\begin{figure}[!t]
\centering
\begin{subfigure}[]{0.49\linewidth}
    \centering
    \includegraphics[width=\textwidth]{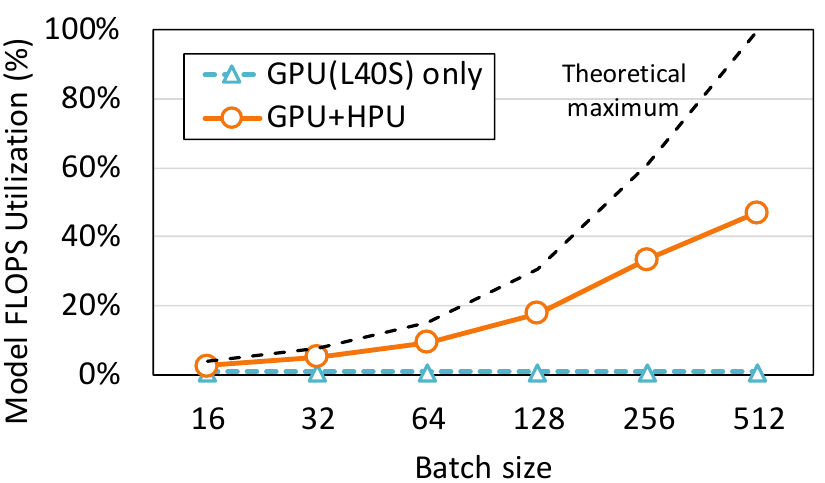}
    \caption{MFU of L40S.}
    \label{fig:eval_mfu_l40s}
\end{subfigure}
\begin{subfigure}[]{0.49\linewidth}
    \centering
    \includegraphics[width=\textwidth]{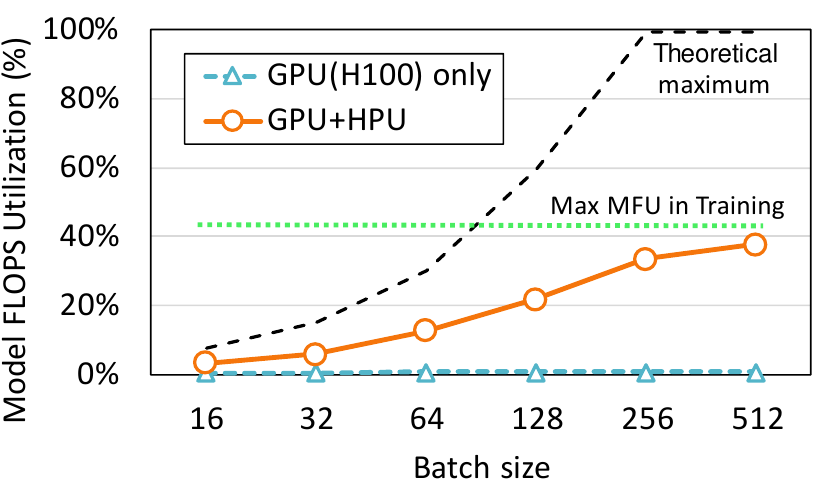}
    \caption{MFU of H100 NVL.}
    \label{fig:eval_mfu_h100}
\end{subfigure}
\caption{Comparison of MFU trends with increasing batch size for GPU-only and GPU-HPU systems during Llama 2 7B inference.}
\label{fig:eval_mfu}
\end{figure}

\subsection{GPU Utilization}
\label{sec:eval_util}
To understand the impact on MFU, we experimented with scaling batch sizes while leaving only compute-bound operations (i.e., linear operations) on the GPU.
Additionally, we projected MFU for the end-to-end scenario of each configuration. The experiment scaled batch sizes from 16 to 512, measuring the utilization of both the L40S and H100 NVL. Fig.~\ref{fig:eval_mfu} presents the results, showing that the GPU-only system, regardless of GPU type, exhibited very low MFU, around 1\%, indicating inefficient use of GPU computing resources. In contrast, the GPU-HPU heterogeneous systems showed increasing MFU as the batch size scaled, reaching a maximum of 44\% for the L40S + HPU prototype and up to 39\% for the H100 NVL with high-performance HPUs. This represents a relatively high value in comparison to the theoretical maximum and closely aligns with the upper bounds typically achievable in LLM training~\cite{mpt_train_bench,semi_h100_cluster}. This implies that a heterogeneous system can improve GPU computational resource utilization by up to 40$\times$. 

\begin{figure}[t]
    \centering
    \includegraphics[width=1\linewidth]{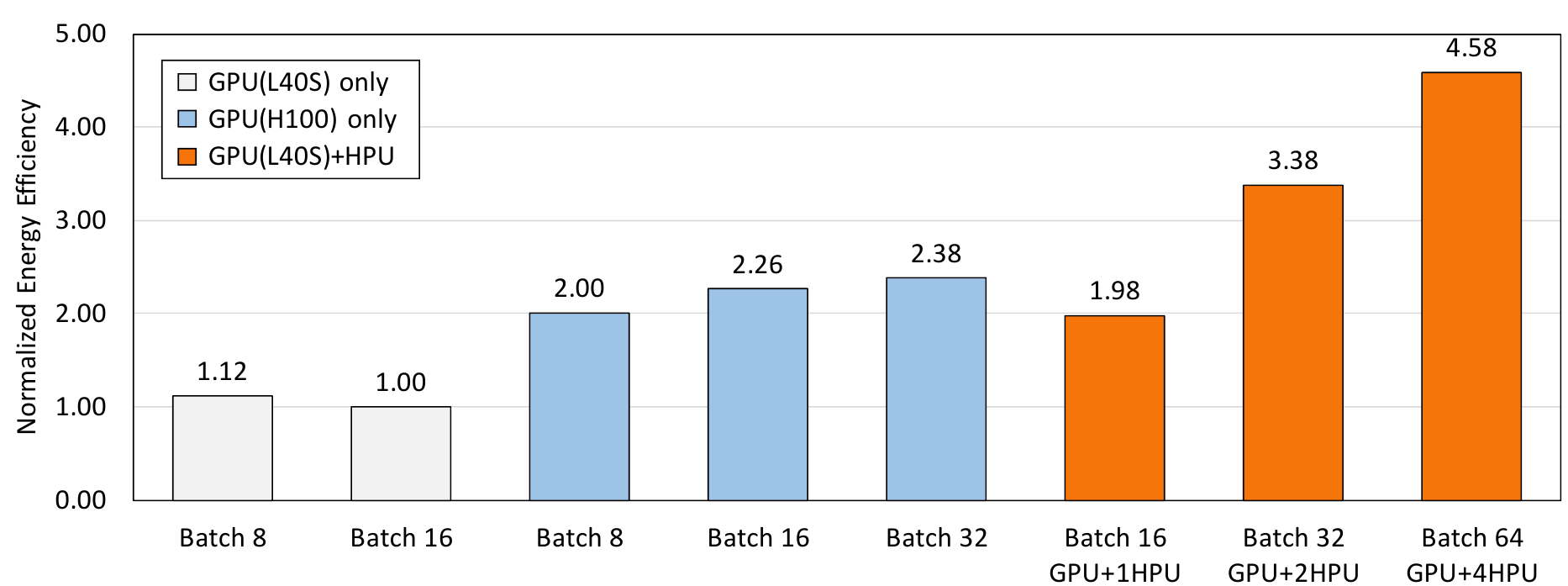}
    \caption{Energy efficiency of GPU system and GPU-HPU heterogeneous system.}
    \label{fig:energy_eff}
\end{figure}

\subsection{Energy Efficiency}
\label{sec:eval_energy}
We measured the power consumption of each system to compare their energy efficiency. Fig.~\ref{fig:energy_eff} shows the normalized energy efficiency in terms of tokens/s/W, which is calculated by dividing throughput by average power consumption. The L40S + HPU prototype system showed up to a 4.58$\times$ improvement in energy efficiency compared to the L40S GPU alone. This indicates that leveraging the HPU prototype in a heterogeneous system is significantly more energy-effective than using a standalone L40S GPU for inference. Furthermore, the L40S + HPU prototype system demonstrated up to a 1.92$\times$ energy efficiency improvement over the H100 NVL-only system, highlighting that a mid-range GPU paired with HPU prototypes can outperform a high-end GPU in terms of inference energy efficiency.

\section{Conclusion}
\label{sec:conclusion}
In this work, we propose the High-Bandwidth Processing Unit (HPU) and GPU-HPU heterogeneous system, an energy-efficient and scalable solution for LLM inference.
The HPU enhances system throughput, energy efficiency, and GPU utilization by offloading memory-bound tasks from GPUs. We developed a real FPGA-based HPU prototype with HBM. 
To the best of our knowledge, the HPU is the first HBM-based co-processor prototype in a heterogeneous system for LLM inference. Experimental results show that the GPU-HPU heterogeneous system improves throughput and energy efficiency by up to 4.1$\times$ and 4.6$\times$, respectively. Also, the mid-range GPU-HPU prototype system achieves a 1.92$\times$ throughput/power gain compared to a high-end GPU system. 

\ifx
\section*{Acknowledgment}
.
\fi

\bibliographystyle{unsrt}
\bibliography{refs}

\end{document}